\documentclass[aps]{revtex4-1}
\usepackage{graphicx}
\usepackage{bm}
\begin{document}

\title{Stream instabilities in relativistically hot plasma}
\author{Rashid Shaisultanov}
\email{rashids@bgu.ac.il}
\affiliation{Physics Department, Ben-Gurion
University, Be'er-Sheva 84105, Israel}
\author{Yuri Lyubarsky}
\email{lyub@bgu.ac.il}
\affiliation{Physics Department, Ben-Gurion
University, Be'er-Sheva 84105, Israel}
\author{David Eichler}
\email{eichler@bgu.ac.il}
\affiliation{Physics Department, Ben-Gurion
University, Be'er-Sheva 84105, Israel}

\begin{abstract}
The instabilities of relativistic ion beams in a relativistically hot electron background are derived for general propagation angles. It is shown that the Weibel instability in the direction perpendicular to the streaming direction is the fastest growing mode,and probably the first to appear, consistent with the aligned filaments that are seen in PIC simulations. Oblique, quasiperpendicular modes grow almost as fast, as the growth rate varies only moderately with angle, and they may distort  or corrugate the filaments after the perpendicular mode saturates.
\end{abstract}
\maketitle
\section{Introduction}

The Weibel instability appears to be important for catalyzing shocks in unmagnetized plasmas. With the discovery of relativistic blast waves from GRB, it has received much attention,\cite{ML99,WA04,Lyub06,Achterberg07,bret09,bret10,almog,waxman}, and appears to be confirmed by simulations. The conventional wisdom is that when two oppositely directed plasma streams collide, the Weibel instability generates small-scale magnetic fields; particle scattering off these magnetic fluctuations provides an isotropization mechanism necessary for the shock transition to form. Particle in cell simulations  \cite{spit2,spit1} show that  a) magnetic filaments grow with a k vector that is perpendicular to the beam direction, b) these filaments saturate at about  10\% of the equipartition level  and are eventually disrupted, and c) the electrons are efficiently heated within the shock structure until their energy becomes comparable to the proton energy.  The full shock transition occurs at a small scale of dozens proton skin depths. The physical reason for the remarkably efficient electron heating has not been fully established yet, however the transfer of energy to the electrons can already be seen in the linear instabilities, which are accompanied by induced electric fields that induce electron countercurrents, as per Lenz's law \cite{blandford,Lyub06,Achterberg07}. It is  necessary to understand the underlying physics before simulations can be scaled and generalized to real astrophysical phenomena.

An important point regarding the very first stages of relaxation is that
the electron streaming is halted easily, whereas protons still plow on through an isotropic electron gas. The newly isotropized electron background effectively suppresses the generation of  magnetic field because of the induced electric field and attendant Lenz  currents opposite to the proton currents \cite{blandford, Lyub06, Achterberg07}. In the case of a Weibel unstable mode with k vector exactly perpendicular to counterstreaming ion beams, the mode has a purely imaginary frequency in the electron frame. The magnitude of the  Lenz currents then decreases with electron temperature. In non-relativistic case, the fraction of electrons that resonate with the purely growing mode decreases with the increasing thermal velocity. In relativistically hot plasma, the electron inertia grows with the plasma temperature. In any case, both the magnitude and the spatial scale of the magnetic fluctuations depend on the "temperature" of the background electrons; the larger the temperature, the stronger the magnetic fluctuations.

As a step towards understanding the filament disruption and electron heating within the shock structure, one should consider the full spectrum of unstable modes in the presence of proton beams within the relativistically hot electron background and calculate how much of the energy is transferred from protons to electrons.
 Two electron heating mechanisms are possible in the linear stages of instability: 1) Lenz currents, mentioned above,  and 2)  Landau damping against hot electrons that move in resonance with the wave, feeling its longitudinal electric field along the direction of the wave's phase velocity vector.  The second happens only when the electrons are already hot enough to be in resonance with a wave of substantial phase velocity. These are the waves that make an oblique angle with the beam direction. The first can occur even for cold electrons, because the phase velocity of the perpendicular Weibel unstable mode can vanish in the frame of the electrons.  Thus the relative importance of the two mechanisms hinges on the role of oblique waves, and the effects of hot electrons on them.

It has been claimed recently that the instability of oblique modes \cite{bret09} and the longitudinal Buneman instability \cite{almog} compete with the Weibel instability
 in the relativistic case.
In this paper we study the full spectrum of the unstable modes for a relativistic beam in relativistically hot ($T\gg m_ec^2)$ electron background, and we show that the electrons Landau damp the oblique modes. The unstable oblique modes then take on more of  Weibel-like character, with corresponding growth rate.

The case of oblique modes has already been studied long ago for cold plasmas. Here, specific properties of relativistic streaming instabilities are dictated by the fact that the response of the relativistic particles to an external perturbation is highly anisotropic. The effective particle mass, namely the ratio of force to acceleration, for longitudinal (with respect to the direction of the  particle motion) perturbations is $m_{\rm eff}=m\gamma^3$ whereas for the transverse perturbations $m_{\rm eff}=m\gamma$; here $\gamma$ is the particle Lorentz factor. For this reason, a  relativistic particle beam excites, in a cold plasma, obliquely propagating Langmuir waves much faster than the waves in the direction of the beam \cite{Bludman1,Bludman2}: the growth rate of the resonant beam instability is $\Gamma\propto m_{\rm eff}^{-1/3}$ so that $\Gamma\propto\gamma^{-1}$ for longitudinal modes
and $\Gamma\propto\gamma^{-1/3}$ for oblique modes, respectively.  Fully electromagnetic analysis of a cold plasma and a cold beam system \cite{Fain70}  { confirmed} that the obliquely propagating unstable modes are practically electrostatic and also revealed an instability of purely transverse perturbations with the growth rate $\Gamma\propto m_{\rm eff}^{-1/2}\propto \gamma^{-1/2}$, which is in fact the relativistic counterpart of the Weibel instability \cite{W59}.

The fact that a relativistic beam is most unstable  to the excitation of oblique waves has important implications for  inertial fusion (e.g. \cite{Bret05}), where the electrons remain subrelativistic. In a relativistically hot plasma, however, electrons can Landau damp oblique waves, which travel at a significant fraction of $c$, and the question should be investigated carefully.



Another interesting aspect of  the simulation results is that current filaments   aligned with the beams are formed initially, but eventually disrupt.  The disruption may be due to non-linear effects, but the filaments may also be disrupted by the eventually emergence of oblique modes, which one suspects would lead to a more complicated magnetic geometry and chaotic particle trajectories.  The randomization of electron trajectories is of course connected to the issue of electron heating, as the rate of heating by electric fields is determined by electric resistivity, which is proportional to the electron scattering rate. So oblique modes may play a role both in electron heating and filament disruption.

In ref. \cite{almog}, the general dispersion equation was solved numerically and the results were reported graphically for a specific sample choices of parameters. Here we use a more analytic approach, in the interest of generality. We present analytic solutions to the general dispersion equation describing a monochromatic beam in the relativistically hot electron background. Generalization to multi-beam systems is straightforward. The mass of the beam particles is not specified here, but is intended in the context of proton beams within the shock structure.

It is shown that the effects of Landau damping are comparable to or less than those of induced transverse electric fields, so that  the general picture that is gaining acceptance - beam instability that is primarily Weibel in character, generating magnetic filaments of negligible phase velocity - more or less holds up even when oblique modes and Landau damping are included in the analysis. On the other hand, the oblique modes, as they become quasiperpendicular, increasingly resemble Weibel unstable electromagnetic modes, and have a growth rate that is a bit slower but  comparable to the latter. So they are not entirely negligible and could eventually lead to a more complex magnetic structure.

We consider here hot electrons but continue to assume cold ion beams. When the ion beams are warm, all instabilities are suppressed, but the Weibel branch is still unstable at sufficiently long wavelengths, where the electrostatic modes are superluminal and stable. So we consider the cold ion beam case the most favorable for the electrostatic instabilities. We will see that even in this case, the Weibel instability is dominated in the relativistically hot plasma.

The paper is organized as follows. In the next section, we present the dispersion equations for the monochrmatic beam in the isotropic electron background. In sect. III we study instability of longitudinal (along the beam direction) modes, the Buneman instability. The purely transverse mode corresponding to the Weibel instability are considered in sect. IV. Stability of oblique modes is addressed in sect. V. Conclusions are presented in sect. VI. In Appendix, we present the permeabilities of the electron plasma with ultrarelativistic Maxwell's distribution.

\section{Dispersion equation}
We consider interaction between a relativistic "monochromatic" beam and the isotropic electron plasma. The stability analysis is reduced to solution of the dispersion equation
\begin{equation}
{\rm Det}|k^{2}\delta _{ij}-k_{i}k_{j}-\omega^{2}\varepsilon_{ij}|=0.
\end{equation}
In this paper, we take the speed of light to be unity.
In the plasma-beam system, one can conveniently separate the dielectric tensor into the plasma and the beam parts, $\varepsilon_{ij}=\varepsilon_{ij}^{\rm pl}+\varepsilon_{ij}^{\rm b}-\delta_{ij}$, and present the plasma dielectric tensor as (e.g. \cite{phkin})
\begin{equation}\label{eqij}
\varepsilon_{ij}^{\rm pl}=\left( \delta _{ij}-\frac{k_{i}k_{j}}{k^{2}}\right)
\varepsilon _{t}+\frac{k_{i}k_{j}}{k^{2}}\varepsilon _{l};
\end{equation}
where $\varepsilon _{t}$ and $\varepsilon _{l}$ are the transverse and longitudinal dielectric permeabilities, respectively. The permeabilities of the Maxwell plasma with highly relativistic temperatures are presented in Appendix. We take a monochromatic beam of particles with the mass $m_b$ and the density $n_b$ moving in $z$ direction with the velocity $v_b$.  Then dielectric tensor of the beam is presented as
 \begin{equation}
 \varepsilon_{ij}^{\rm b}=\delta _{ij}\left(1-\frac{\omega _{b}^{2}}{\omega^{2}\gamma_b}\right)-\frac{\omega _{b}^{2}}{\omega^{2}\gamma_b}\left(\frac{k_{i}v_{bj}+k_{j}v_{bi}}{\omega-k_{z}v_b}+\frac{k^{2}-\omega^{2}}{\left(\omega-k_{z}v \right)^{2}}v_{bi}v_{bj}\right);
 \end{equation}
where
\begin{equation}
\omega_{b}^{2}=4\pi n_{b} e^{2}/m_{b}.
\end{equation}

We choose the coordinate system such that $k_y=0$. Then the dispersion equation for the plasma-beam system is reduced, after some algebra, to the form
\begin{eqnarray}\label{disp}
\left[ \varepsilon _{l}-\frac{\omega _{b}^{2}}{\gamma_b^{3}\left( \omega
-k_{z}v_b\right) ^{2}}\right] \left[ k^{2}-\omega ^{2}\varepsilon _{t}+\frac{%
\omega _{b}^{2}}{\gamma_b}\right] =\frac{\omega _{b}^{2}k_{x}^{2}v^{2}_b}{%
\gamma \left( \omega -k_{z}v_b\right) ^{2}}\left[ 1-\varepsilon _{l}\left(1-\frac{\omega^2}{k^2}\right)-\frac{\omega^2}{k^2}\varepsilon _{t}\right]
\end{eqnarray}
If the plasma is cold,
$\varepsilon _{t}=\varepsilon _{l}=1-\omega_{p}^{2}/\omega^{2}$, $\omega_p^2=4\pi e^2n/m_e $, equation (\ref{disp}) reproduces the result of \cite{Fain70}.

For all practical purposes we can safely assume that
\begin{equation}
\frac{\omega_b}{\sqrt\gamma_b}\ll\frac{\omega_p}{\sqrt{\gamma_T}}.
\end{equation}
For electron beams, this implies $n_b/\gamma_b\ll n_p/\gamma_T$ whereas for proton beams,
which is of special interest in the context of relativistic shocks, this condition is fulfilled even for $n_b\sim n_p$, $\gamma_b\sim\gamma_T$.

\section{Buneman instability}
First we consider instability of the wave propagating along the beam, $\vec{k}=(0,0,k)$. In this case the dispersion equation (\ref{disp}) is reduced to
 \begin{equation}
 \varepsilon _{l}=\frac{\omega _{b}^{2}}{\gamma_b^{3}\left( \omega
-kv_b\right) ^{2}}.
 \end{equation}
The instability occurs due to the resonance
of particles with waves, so that one can write $\omega=kv_b+\delta\omega$; $\delta\omega\ll\omega$. For a highly relativistic beam, $\omega\approx k$, therefore we have to use equation (\ref{eps_l1}) for $\epsilon_l$.  Then the dispersion equation is reduced to the cubic equation
 \begin{equation}
\delta\omega^2\left[\delta\omega+\xi(k-k_0)-\frac{k_0}{2\gamma_b^2}\right]=
\frac{\omega^2_b\omega_0^3}{12\omega_p^2\gamma_T\gamma^3_b}.
 \end{equation}
The maximal growth rate is achieved at $k=k_0[1+1/(2\xi\gamma^2_b)]$ and is equal to
\begin{equation}\label{}
    \Gamma_{\rm Buneman}=\frac{3^{1/6}}{2^{5/3}}\left( \frac{\omega_{b}}{\omega_{p}}\right )^{\frac{2}{3}}\frac{\omega_{0}}{\gamma_{b}\gamma_T^{1/3}}=
    \frac{3^{1/6}}{2^{5/3}}\frac{\omega_{b}^{2/3}\omega_{p}^{1/3}\sqrt{\ln2\gamma_T}}{\gamma_{b}\gamma_T^{5/6}}.
\end{equation}

The Buneman regime occurs at the condition $\gamma_b>\gamma_T$. When the beam velocity is less than the plasma thermal velocities, $\gamma_b\ll\gamma_T$, the beam is still unstable, but the growth rate decreases. According to ref. \cite{Lom79}, the maximal growth rate in this case is estimated as
\begin{equation}
\Gamma\approx \frac{\omega_b\sqrt{2\ln\gamma_b}}{\gamma_b^{3/2}}
\end{equation}

\section{Weibel instability}
Now let us consider instability of transverse waves, $\vec{k}=(k,0,0)$. In this case one can neglect the terms with $\omega_b$ in the left-hand side of the dispersion equation (\ref{disp}). The right-hand side becomes comparable with the left-hand side only at $\omega \ll k$.
In this limit, the plasma dielectric permeabilities (\ref{silin_l}) and (\ref{silin_t}) are reduced to
\begin{equation}
 \varepsilon _{l}=1+\frac{\omega_{p}^{2}}{\gamma_Tk^2} ;\;\;\;  \varepsilon _{t}=1-\frac{\omega_{p}^{2}}{\gamma_Tk^2}+i\frac{\pi\omega_{p}^{2}}{4\gamma_T\omega k}.
 \end{equation}
Now the dispersion equation (\ref{disp}) is reduced to the cubic equation
 \begin{equation}
 \left(1+\frac{\omega_{p}^{2}}{\gamma_Tk^2}\right)\left(k^2- i\frac{\pi\omega_{p}^{2}\omega}{4  \gamma_Tk}\right)=-\frac{\omega_{b}^2 \omega_{p}^{2}}{\gamma_T\gamma_b \omega^2}
 \label{weib_disp}\end{equation}
 \begin{figure*}
  \includegraphics[width=1.0 \textwidth]{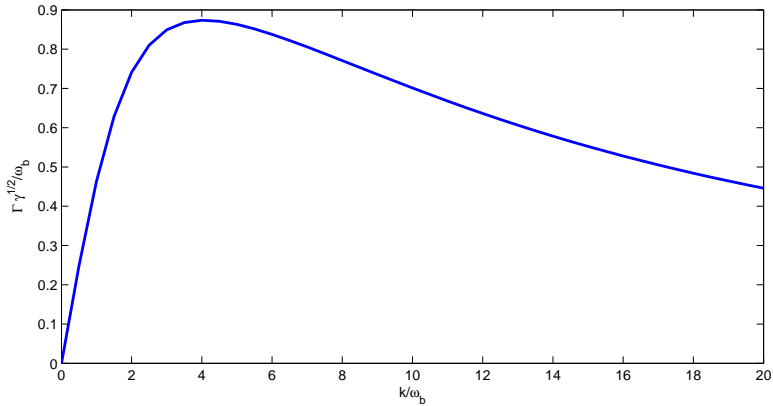}\\
\caption{Growth rate of the Weibel instability of the proton beam at $\gamma_{b}=100$, $\gamma_T=20$ and equal densities $n_{b}=n$}\label{Fig2}
\end{figure*}
The unstable solutions to this equation are purely imaginary, which implies aperiodic instability. The growth rate has a broad maximum (see Fig.\ref{Fig2})
 \begin{equation}
\Gamma_{\rm Weibel}\approx\frac{\omega_b}{\sqrt{\gamma_b}}
 \label{weibel}\end{equation}
 at
 \begin{equation}
k\sim  \left(\frac{\omega_p^2\omega_b}{\gamma_T\gamma_b^{1/2}}\right)^{1/3}.
 \end{equation}
In the long and small wavelength limits the growth rate is given by:
\begin{eqnarray}
    \Gamma=\left(\frac{4\omega_b^2\gamma_T}{\pi\omega_p^2\gamma_b}\right)^{1/3}k;
    \qquad k\ll \left(\frac{\omega_p^2\omega_b}{\gamma_T\gamma_b^{1/2}}\right)^{1/3};\\
    \Gamma=\frac{\omega_{b}}{\sqrt{\gamma}}\frac{\omega_{p}}{(\gamma_Tk^{2}+\omega_{p}^{2})^{1/2}};
    \qquad k\gg \left(\frac{\omega_p^2\omega_b}{\gamma_T\gamma_b^{1/2}}\right)^{1/3};
\end{eqnarray}

Comparing the Buneman and the Weibel instabilities, one finds that the ratio of the growth rates very weakly depends on the mass and density ratios of the beam and plasma particles:
\begin{equation}
\frac{\Gamma_{\rm Buneman}}{\Gamma_{\rm Weibel}}=
\frac{3^{1/6}}{2^{5/3}}\left(\frac{\omega_{p}}{\omega_b}\right)^{1/3}
\frac{\sqrt{\ln2\gamma_T}}{\gamma_{b}^{1/2}\gamma_T^{5/6}}=
\frac{3^{1/6}}{2^{5/3}}\left(\frac{m_bn_{p}}{m_en_b}\right)^{1/6}
\frac{\sqrt{\ln2\gamma_T}}{\gamma_{b}^{1/2}\gamma_T^{5/6}}.
\end{equation}
Therefore in the relativistic case, $\gamma_T,\gamma_b\gg 1$, the Weibel instability dominates the Buneman one even for the proton beam unless the density of the beam is extraordinarily small.

\section{Instability of oblique modes}
Now let us consider oblique modes. In this case one can also neglect the terms with $\omega_b$ in the left-hand side of the dispersion equation (\ref{disp}). The right-hand side becomes comparable with the left-hand side only for modes in resonance with the beam, i.e. if $\omega\approx k_z=k\cos\theta$. Therefore one can substitute $\omega=k\cos\theta$ into all the terms with the exception of the resonance denominator, which yields
 \begin{equation}
\varepsilon_l(1-\cos^2\theta\varepsilon_t)=
\frac{\omega _{b}^{2}\sin^{2}\theta }{%
\gamma \left( \omega -kv_b\cos\theta\right) ^{2}}\left[ 1-\sin^2\theta\varepsilon _{l}-\cos^2\theta\varepsilon _{t}\right];
 \end{equation}
where one can use the permeabilities in the form (\ref{silin_l}) and (\ref{silin_t}) with $\omega=k\cos\theta$:
\begin{eqnarray}
\varepsilon_{l}=1+\frac{\omega_p^{2}}{k^{2}\gamma_T }\left(1-\cos\theta\ln\cot\frac{\theta}2+
\frac{\pi}2i\cos\theta\right)\\
\varepsilon_{t}=1-\frac{\omega_p^{2}}{2 k^2\gamma_T\cos\theta} \left(\sin^2\theta\ln\cot\frac{\theta}2 +\cos\theta-\frac{\pi}2 i\sin^2\theta\right).
\end{eqnarray}
Note that these expression diverge at $\theta=0$ and $\theta=\pi/2$; therefore these specific cases should be considered separately; this has been done in sect. III and IV.

\begin{figure*}
\includegraphics[width=1.0 \textwidth]{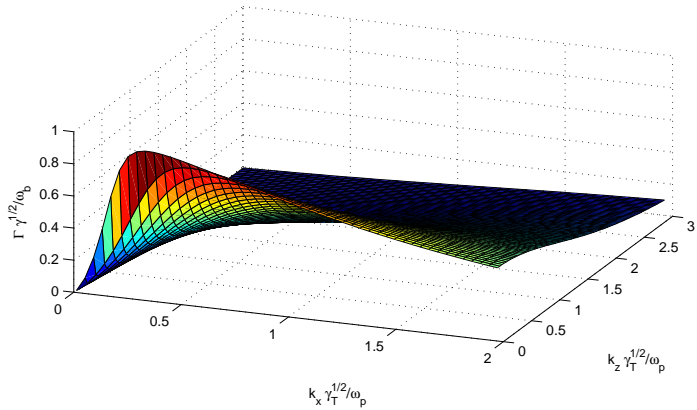}\\
  \caption{Growth rate of the oblique modes as a function of  $k_{x},k_{z}$. The beam is in the $z$ direction. 
  }\label{Fig3}
\end{figure*}

Now the solution to the dispersion equation is written as
 \begin{eqnarray}
\omega=k\cos\theta\pm\frac{\omega_b}{\sqrt{\gamma_b}}\Phi\left(\frac{\omega_p^2}{\gamma_Tk^2},\theta\right);\\
 \Phi=\sin\theta\sqrt{\frac{1-\sin^2\theta\varepsilon _{l}-\cos^2\theta\varepsilon _{t}}{\varepsilon_l(1-\cos^2\theta\varepsilon_t)}}.
 \end{eqnarray}
The growth rate of the instability is presented as
\begin{equation}
\Gamma=\frac{\omega_b}{\sqrt{\gamma_b}}\,{\rm Im}\,\Phi\left(\frac{\omega_p^2}{\gamma_Tk^2},\theta\right).
\end{equation}
One sees that the dependence on the beam parameters is split from the dependence on the plasma parameters therefore we can find the universal dependence of the growth rate on the parameters by normalizing
$\Gamma$ by $\omega_b/\sqrt{\gamma_b}$ and $k$ by $\omega_p/\sqrt{\gamma_T}$. This dependence is presented in
Fig. (\ref{Fig3}).
\begin{figure*}
\includegraphics[width=1.0 \textwidth]{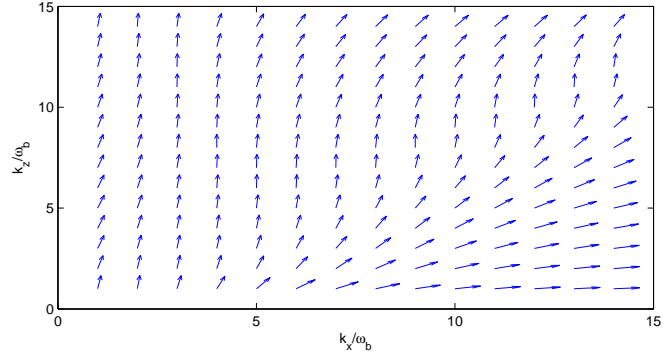}\\
  \caption{The direction of the electric field in the oblique modes; $\gamma_{b}=100$, $\gamma_T=20$, $n=2n_{b}$. Note that as $k$ decreases, the modes become more electromagnetic in character.} \label{Fig6}
\end{figure*}
The polarization of the unstable modes is hsown in Fig. (\ref{Fig6}).
In Fig. (\ref{Fig4}), the growth rate is presented as a function of k at a fixed angle $\theta=\pi /4$.
\begin{figure*}
\includegraphics[width=1.0 \textwidth]{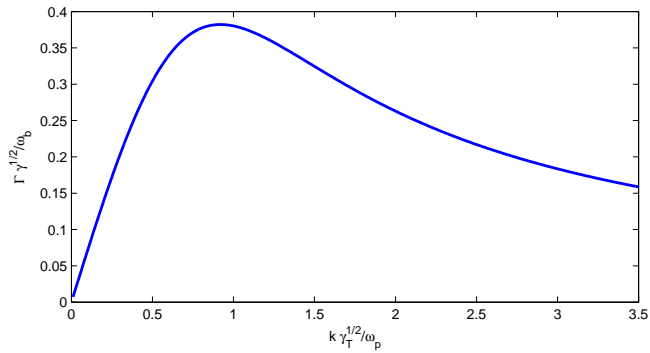}\\
  \caption{Growth rate of the oblique mode as a function of k at $\theta=\pi /4$} \label{Fig4}
\end{figure*}
In Fig. (\ref{Fig5}), we plot the maximal growth rate as a function of the angle between the wave and the beam. One sees that the largest growth rate is achieved for the transverse mode, i.e. for the Weibel instability.
\begin{figure*}
\includegraphics[width=1.0 \textwidth]{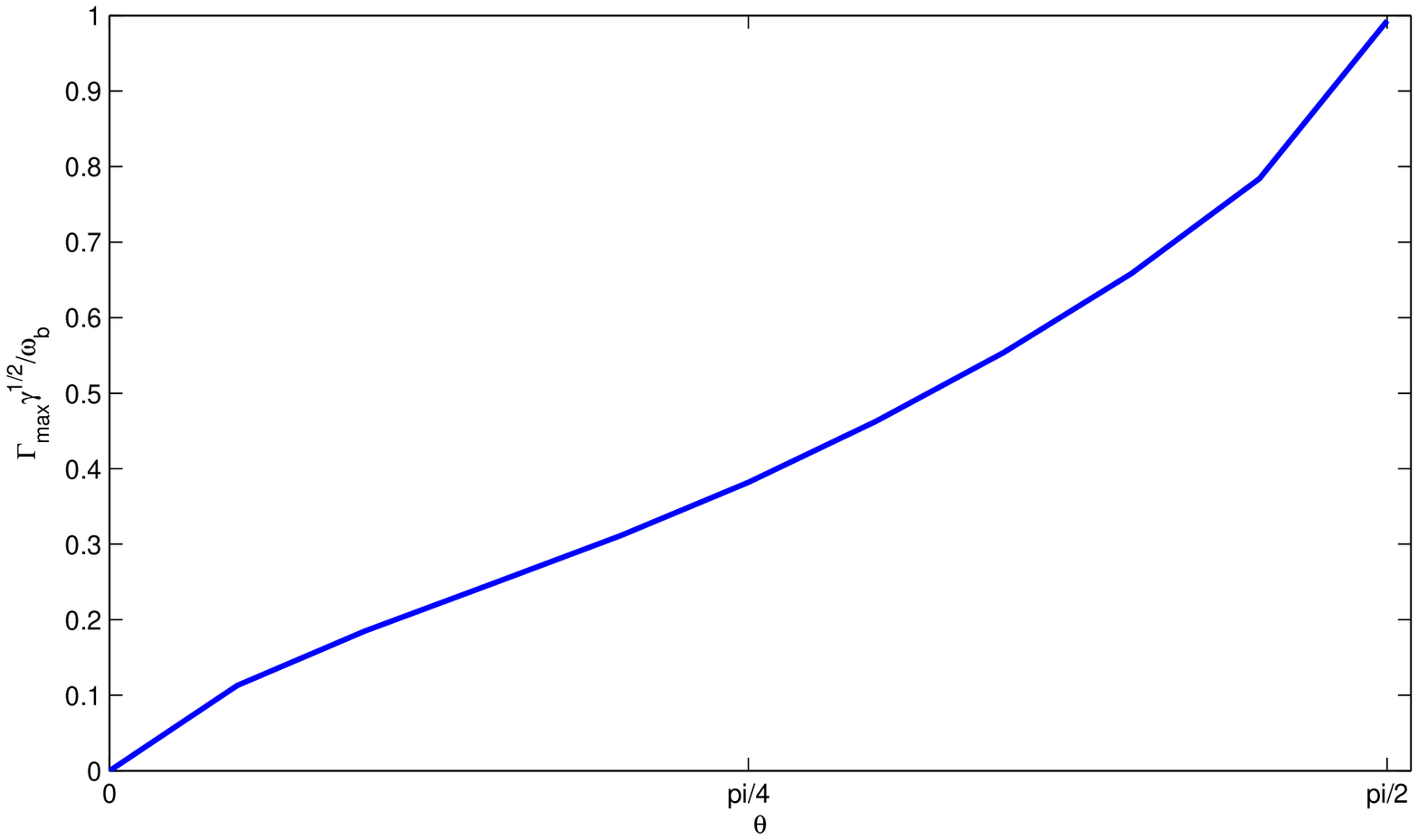}\\
  \caption{The maximal growth rate as a function of $\theta$} \label{Fig5}
\end{figure*}

\section{Discussion and Conclusions}
We found the full set of unstable modes for the monochromatic relativistic beam propagating through relativistically hot electron background.
The results are easily generalized to the multi-stream case. The oblique and longitudinal modes are resonant, therefore each beam excites the appropriate wave independently of others. In the case of the Weibel instability, one has to substitute, for the term $\omega^2_b/\gamma_b$ in the right-hand side of the dispersion equation (\ref{weib_disp}),the  sum  over all the beams  $\sum_j\omega^2_{b,j}/\gamma_{b,j}$; this yields the same substitution in the expressions for the growth rate (see also \cite{Lyub06}).

Our analysis confirms that the strongest is the Weibel instability with wavevector perpendicular to the beam direction, with the growth rate given by equation (\ref{weibel}). The reason  the oblique modes are suppressed, relative to the case of a relativistic beam in a non-relativistic plasma, is that the resonance waves are subluminal, $\omega/k=v_b\cos\theta$, therefore in a relativistically hot plasma, they are suppressed by the Landau damping. The reason, on the other hand, the oblique modes grow more slowly than the perpendicular Weibel instability, which is also suppressed by the electrons, is that the projection of the beam anisotropy on the k vector is reduced by sin$\theta$, so such suppression is modest when $\theta$ is close to $\pi/2$.

Support from the Israel Science
Foundation,  the Israel-U.S. Binational Science Foundation, and the
Joan and Robert Arnow Chair of Theoretical Astrophysics  is
gratefully acknowledged.

\section*{Appendix. Dielectric permeabilities of relativistically hot plasma.}

Generally the longitudinal and transverse permeabilities are written as (e.g. \cite{phkin})
\begin{eqnarray}
\varepsilon_{l}^{\rm pl}=1
+\frac{4\pi e^2}{k^2}\int \mathbf{k\cdot}\frac{\partial f(\mathbf{p})}{\partial \mathbf{p}}\frac{d^{3}p}{\omega+i0-\mathbf{k\cdot v}}
\label{eps_l}\\
\varepsilon_{t}^{\rm pl}=1+\frac{4\pi e^2}{\omega^2}\int\left(\mathbf{v}-\frac{\mathbf{k\cdot v}}{k^2}\mathbf{k}\right)\cdot\frac{\partial f(\mathbf{p})}{\partial \mathbf{p}}\frac{d^{3}p}{\omega+i0-\mathbf{k\cdot v}};\label{eps_t}
\end{eqnarray}
We assume that the plasma electrons have Maxwell's distribution
\begin{equation}
f(\mathbf{p})=\frac n{8\pi (m_e\gamma_T)^{3}}\exp\left(-\frac{\gamma}{\gamma_T}\right).
\end{equation}
with ultrarelativistic temperatures, $\gamma_T\gg 1$.
Integration over the angles in Eqs. (\ref{eps_l}) and (\ref{eps_t}) is performed straightforwardly. Integration over $p$ could be performed if one substitutes $v=1$; this yields
\cite{Silin60} (see also \cite{phkin})
\begin{eqnarray}
\varepsilon_{l}^{\rm pl}=1+\frac{\omega_p^{2}}{k^{2}\gamma_T }\left[1+\frac{\omega}{2k}\ln\left(\frac{\omega+i0-k}{\omega+k}\right)\right]
\label{silin_l}\\
\varepsilon_{t}^{\rm pl}=1+\frac{\omega_p^{2}}{4\omega k\gamma_T }\left[\left(1-\frac{\omega^{2}}{k^{2}}\right)\ln\left(\frac{\omega+i0-k}{\omega+k}\right)-\frac{2 \omega}{k}\right];
\label{silin_t}\end{eqnarray}
where
 \begin{equation}
 \omega_p^2=\frac{4\pi e^2n}{m_e}.
 \end{equation}

\begin{figure*}
  \includegraphics[width=1.0 \textwidth]{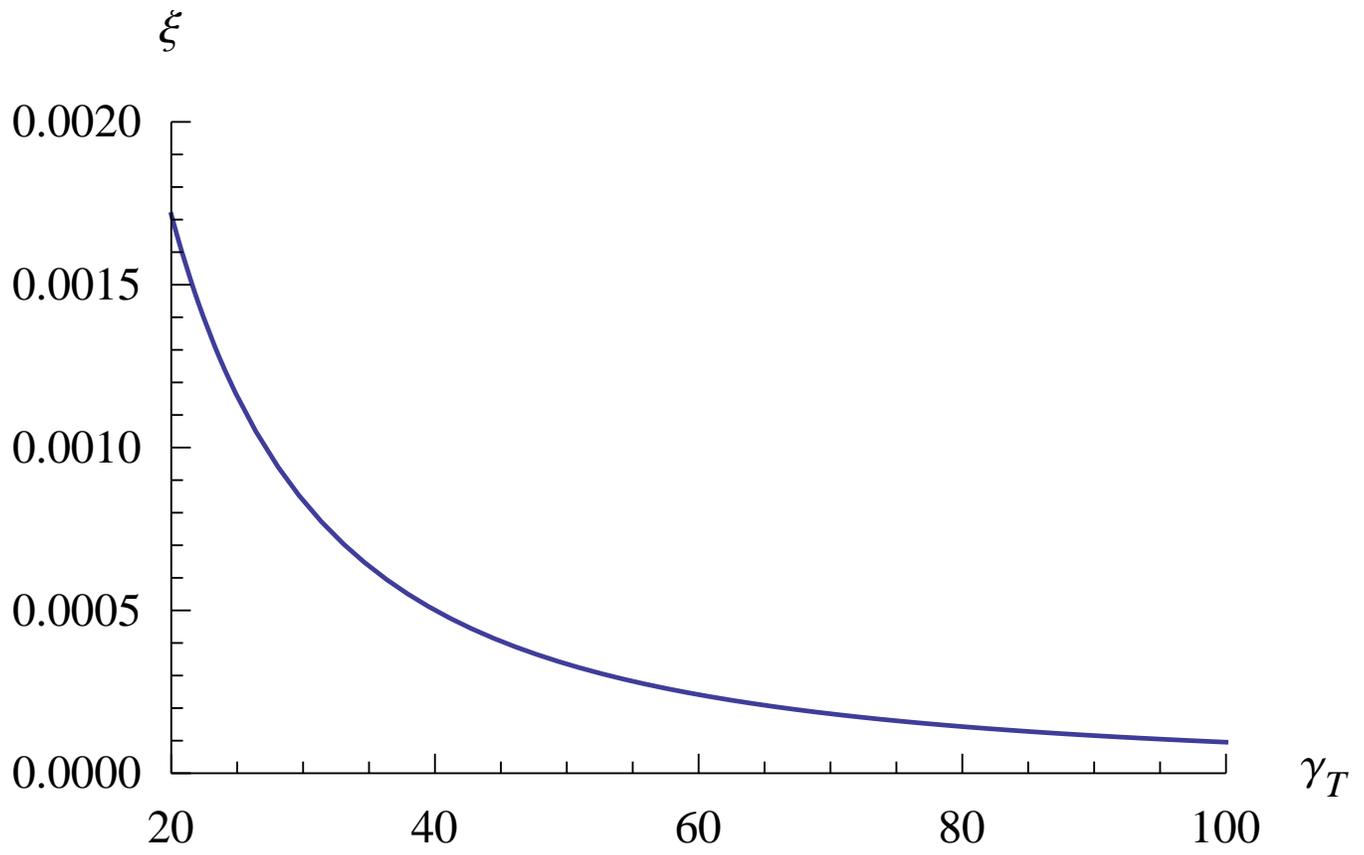}\\
  \caption{The parameter $\xi$ (see Eq. (\ref{omega01})) as a function of the plasma temperature}\label{Fig1}
\end{figure*}

Note that the expression (\ref{silin_l}) diverges at $\omega\to k$. In this case, one has to take into account small deviations of $v$ from the speed of light in denominators $\omega-\mathbf{k\cdot v}$ in the exact formula (\ref{eps_l}). Only then one can find correctly describe the longitudinal waves with the phase velocity close to the speed of light and in particular,   subluminal modes crucially important for the beam instability  \cite{Lom79}. Following \cite{Lom79} we first find the longitudinal wave with the phase velocity equal to speed of light $\omega_0/k_0=1$. In this case the dispersion equation, $\varepsilon_l=0$, with the exact permeability (\ref{eps_l}) yields
\begin{equation}
\omega_0^2=k_{0}^2=\frac{\omega _{p}^{2}}{\gamma_T}\ln 2\gamma_T.
\end{equation}
Now we could expand $\varepsilon_l$ in small $\omega-\omega_0$, $k-k_0$ to yield the longitudinal permeability in the vicinity of the point $\omega_0,k_0$:
 \begin{equation}
\varepsilon_l^{\rm pl}(\omega,k)=\left(\frac{\partial \varepsilon _{l}^{\rm pl} }{\partial \omega}\right)_{0}(\omega-\omega_0)+\left(\frac{\partial \varepsilon _{l}^{\rm pl} }{\partial k}\right)_{0}(k-k_0).
 \label{eps_l1}\end{equation}
The zero subscript means that the derivatives are taken at $k=k_{0}$, $\omega=\omega_{0}$.
Straightforward calculation yields
 \begin{equation}
\left(\frac{\partial \varepsilon _{l}^{\rm pl}}{\partial \omega}\right)_{0}=\frac{12 \omega _{p}^{2} \gamma_T}{\omega_{0}^3};
 \label{omega0}\end{equation}
and $\left(\partial \varepsilon _{l}^{\rm pl}/\partial k \right)_{0}$ can be represented as
 \begin{equation}
 \left(\frac{\partial \varepsilon _{l}^{\rm pl}}{\partial k}\right)_{0}=-\left(1-\xi\right)\left(\frac{\partial \varepsilon _{l}^{\rm pl}}{\partial \omega}\right)_{0};
 \label{omega01}\end{equation}
where $\xi$ is some function of temperature, it is small as $\gamma_T^{-2}$ (see Fig. \ref{Fig1}).
Since $\left(\frac{\partial \varepsilon _{l}^{\rm pl}}{\partial \omega}\right)_{0}>
 \left|\left(\frac{\partial \varepsilon _{l}^{\rm pl}}{\partial k}\right)_{0}\right|$, the waves with $k>k_0$
are subluminal; these waves could be resonantly excited by a relativistic particle beam, see sect. III.

\end{document}